\documentclass[reprint, superscriptaddress, secnumarabic, amssymb, nobibnotes, aps, prl]{revtex4-1}

\usepackage{graphicx}
\usepackage{epstopdf}
\usepackage[T1]{fontenc}
\usepackage[latin9]{inputenc}
\usepackage{amsbsy}
\usepackage{gensymb}
\usepackage[T1]{fontenc}
\usepackage[latin9]{inputenc}
\usepackage{amsmath}
\usepackage{amssymb}
\usepackage{bbm}
\usepackage{braket}
\usepackage{xcolor}
\allowdisplaybreaks
\usepackage{graphicx}
\usepackage[colorlinks=true]{hyperref}  
\hypersetup{
    bookmarks=true,         
    unicode=false,          
    pdftoolbar=true,        
    pdfmenubar=true,        
    pdffitwindow=false,     
    pdfstartview={FitH},    
    pdftitle={x},    
    pdfauthor={SM},     
    pdfsubject={},   
    pdfcreator={},   
    pdfproducer={}, 
    pdfkeywords={} {} {}, 
    pdfnewwindow=true,      
    colorlinks=true,       
    linkcolor=blue, 
    citecolor=blue,        
    filecolor=magenta,      
    urlcolor=blue           
} 
\usepackage[normalem]{ulem}

\renewcommand{\approx}{\simeq}

\begin{document}

\preprint{AIP/123-QED}

\title{Normal state and superconducting state properties of high entropy Ta$_{0.2}$Nb$_{0.2}$V$_{0.2}$Ti$_{0.2}$X$_{0.2}$ (X = Zr and Hf)}

\author{Nikita Sharma}
\affiliation{Department of Physics and Material Science, Thapar Institute of Engineering and Technology, Patiala 147004, India}
\author{J. Link}
\affiliation{National Institute of Chemical Physics and Biophysics, 12618 Tallinn, Estonia}
\author{Kuldeep Kargeti}
\affiliation{Department of Physics, Bennett University, Greater Noida-201310, Uttar Pradesh, India}
\author{Neha Sharma}
\affiliation{Department of Physics and Material Science, Thapar Institute of Engineering and Technology, Patiala 147004, India}
\author{I. Heinmaa}
\affiliation{National Institute of Chemical Physics and Biophysics, 12618 Tallinn, Estonia}
\author{S. K. Panda}
\affiliation{Department of Physics, Bennett University, Greater Noida-201310, Uttar Pradesh, India}
\author{R. Stern}
\affiliation{National Institute of Chemical Physics and Biophysics, 12618 Tallinn, Estonia}
\author{Tirthankar Chakraborty}
\email[]{tirthankar@thapar.edu}
\affiliation{Department of Physics and Material Science, Thapar Institute of Engineering and Technology, Patiala 147004, India}
\author{Tanmoy Chakrabarty}
\email[]{tanmoy.chakrabarty@krea.edu.in}
\affiliation{National Institute of Chemical Physics and Biophysics, 12618 Tallinn, Estonia}
\affiliation{Division of Sciences, Krea University, Sri City, Andhra Pradesh 517646, India}
\author{Sourav Marik}
\email[]{soumarik@thapar.edu}
\affiliation{Department of Physics and Material Science, Thapar Institute of Engineering and Technology, Patiala 147004, India}

\date{\today}
\begin{abstract}
\begin{flushleft}
\end{flushleft}
High entropy alloy superconductors represent a unique blend of advanced material systems and quantum physics, offering significant potential for advancing superconducting technologies. In this study, we report a detailed theoretical and experimental investigation of high entropy alloy superconductors Ta$_{0.2}$Nb$_{0.2}$V$_{0.2}$Ti$_{0.2}$X$_{0.2}$ (X = Zr and Hf). Our study unveils that both the materials crystallize in a body-centered cubic structure (space group: I m -3 m) and exhibit bulk superconductivity with a superconducting onset temperature of (T$_C^{onset}$) of 5 K for X = Hf and 6.19 K for X = Zr sample. Our detailed analysis, including magnetization, resistivity, heat capacity measurements, and density functional theory (DFT) calculations indicates moderately coupled isotropic s-wave superconductivity in these materials. Our DFT results find significant spectral weight at the Fermi energy and phonon spectra is free of imaginary modes, confirming the dynamical stability and metallic nature of these alloys. Remarkably, we have observed a high upper critical field (H$_{C2}(0)$) surpassing the Pauli paramagnetic limit for the X = Hf sample and explained it on the basis of the increased spin-orbit coupling in the structure.  Ta$_{0.2}$Nb$_{0.2}$V$_{0.2}$Ti$_{0.2}$Zr$_{0.2}$, on the other hand, shows a conventional H$_{C2}$ behaviour. With the dynamical stability of these alloys, excellent normal state metallic nature, high micro-hardness, and high upper critical field, these samples emerge as potential candidates for future applications in superconducting devices.

\end{abstract}
\maketitle

\section{Introduction}
Quantum materials have recently garnered significant research interest due to their potential to revolutionize a wide range of technological fields, including quantum computing, energy storage, and quantum sensing \cite{1}. Superconductors, in particular, play a critical role among quantum materials because of their unique ability to maintain quantum coherence and facilitate the creation of qubits, the fundamental units of quantum information \cite{2}.

The recent discovery of quantum phenomena like superconductivity in highly disordered, compositionally complex high-entropy alloys (HEAs) has sparked significant research interest, as these materials offer promising new avenues for developing advanced superconducting materials \cite{3,4,5,6,7}. Unlike traditional alloys, which typically contain one or two principal elements, HEAs consist of five or more elements mixed in near-equiatomic ratios \cite{8,9}. Thermodynamically, the formation of single-phase solid solutions in HEAs is driven by the dominance of the entropy of mixing (\(\Delta S_{\text{mix}}\)) in the Gibbs free energy equation: \[\Delta G_{\text{mix}} = \Delta H_{\text{mix}} - T \Delta S_{\text{mix}}\]
where T = temperature and $\Delta H_{\text{mix}}$ = mixing enthalpy. This high configurational entropy stabilizes disordered solid solutions in simple crystal structures, for instance, in face-centered cubic (fcc), body-centered cubic (bcc), and hexagonal close-packed systems (hcp), akin to pure metals. 

The large atomic disorder, resulting from multiple elements with slightly different atomic sizes, is believed to play a part in their novel physical and exceptional mechanical properties \cite{10}. For instance, HEAs show high fracture toughness \cite{11}, hardness \cite{10}, strength, and superior corrosion resistance \cite{12}. These properties make them valuable for various applications, including structural materials, applications in aerospace,  magnetic cooling, energy storage, and superconducting magnets, radiation protection, and bio-compatible advanced materials.

Superconducting HEAs, first observed in a TaNbHfZrTi alloy \cite{3} with a transition temperature of 7.3 K, combine the robust mechanical properties of HEAs with the fascinating quantum phenomenon of superconductivity. The HEA and medium entropy alloy (MEA) superconductors exhibit intriguing superconducting characteristics, such as retention of superconductivity under extremely high pressure \cite{13, 14}, enhanced upper critical fields in a few materials \cite{15, 16}, high critical current densities \cite{17, 18, 19}, Debye temperature within the range typical for elemental superconductors \cite{13, 18, 20, 21} and significant broadening in specific heat jumps. Also, the ability to convert these materials into thin films and their robust fracture strength at cryogenic temperatures make them promising candidates for superconducting devices even in extreme conditions. For instance, TaNbHfZrTi films exhibit superconductivity that is over 1000 times more resistant to displacement damage than other superconductors \cite{18}. This resilience highlights the potential of HEA superconductors for application in extreme conditions, including aerospace applications, nuclear fusion reactors, and high-field superconducting magnets.

Despite their highly disordered nature, high-entropy alloy superconductors exhibit phonon-mediated superconductivity, similar to the observed superconductivity in binary and ternary superconducting alloys, even though regular phonon modes are generally less likely to occur in such disordered systems. This makes superconducting HEAs particularly intriguing, as they provide a unique opportunity to investigate the intricate interplay between disorder and superconductivity. So far, the understanding of BCS superconductivity in HEAs without having the conventional phonon modes remains a significant challenge \cite{22}, mainly due to the limited research in this field. Identifying and characterizing both the normal and superconducting properties of HEAs is crucial for advancing our knowledge of these complex, disordered systems. At the same time, the multi-component nature of HEAs not only allows for tunability and design of quantum materials but also provides insights into the fundamental aspects of superconductivity in disordered systems. 

Herein, we report a thorough experimental and theoretical study of the normal state and superconducting properties of two moderately dense equiatomic HEA superconductors with compositions Ta$_{0.2}$Nb$_{0.2}$V$_{0.2}$Ti$_{0.2}$X$_{0.2}$ (X = Hf and Zr). The inherent maximal disorder in this equiatomic HEA presents a valuable opportunity to study the superconductivity in a high entropic and high disorder system. Our detailed specific heat and resistivity measurements highlight the possibility of unconventional superconductivity in Ta$_{0.2}$Nb$_{0.2}$V$_{0.2}$Ti$_{0.2}$Hf$_{0.2}$. At the same time, a high micro-hardness is observed in these materials. Besides high hardness, the normal state and superconducting state properties of these two materials are presented and compared here.

\section{Methodology}
\textbf{Experimental Details:} The polycrystalline samples of Ta$_{0.2}$Nb$_{0.2}$V$_{0.2}$Ti$_{0.2}$X$_{0.2}$ with X = Zr and Hf were synthesized via the arc melting method. High-purity elemental precursors were precisely weighed according to their stoichiometric ratios and melted several times under a high-purity argon atmosphere. A minimal weight loss was recorded during the synthesis process. The phase purity of the synthesized sample was confirmed by room-temperature powder X-ray diffraction method (XRD) using a Rigaku diffractometer with Cu-K${\alpha 1}$ radiation ($\lambda = 1.5406$ \text{\AA}). Energy Dispersive X-ray Spectroscopy (EDS) mappings were obtained using a BRUKER XFlash 6160 system, while microstructure analysis was performed with a ZEISS GEMINI field emission scanning electron microscope (FE-SEM). Vickers microhardness was assessed on the polished surfaces using an OMNITECH MVH-1C microhardness tester, with a 500 g load applied for 20 seconds. A vibrating Sample Magnetometer (VSM) attached to the Quantum Design Physical Property Measurement System (QD - PPMS) is used for all the Magnetization measurements. Field-cooled (FC) and Zero field-cooled (ZFC) modes were used to record the magnetic susceptibility data of the sample.  Furthermore, the resistivity ($\rho$) and specific heat capacity (C) measurements were performed (function of temperature and magnetic field) using QD - PPMS. 

\textbf{Computational Details:} To model the atomic scale structure of the disordered high entropy alloy (HEA) Ta$_{0.2}$Nb$_{0.2}$V$_{0.2}$Ti$_{0.2}$X$_{0.2}$ (X = Hf and Zr), we employed the special quasi-random structure (SQS) approach, which provides a reliable representation of the disordered solid solution. The SQS was generated using the Monte Carlo-based simulated annealing algorithm implemented in the mcsqs module of the Alloy Theoretic Automated Toolkit (ATAT)~\cite{23, 24, 25}. This method optimizes both the cell shape and atomic site occupancies to match as closely as possible the pair and higher-order correlation functions of a fully random alloy. Specifically, we constructed a 10-atoms SQS that accurately reproduces the pair correlation functions up to the first nearest-neighbor shell, ensuring an unbiased representation of the disordered nature of the HEA. Using this cell, we carried out electronic structure calculations by employing density functional theory (DFT) approaches as implemented in Quantum ESPRESSO~\cite{26}. The exchange-correlation energy was described using the Perdew-Burke-Ernzerhof (PBE) functional within the generalized gradient approximation (GGA)~\cite{27}. A plane-wave energy cutoff of 50 Ry was employed to ensure numerical accuracy. The Brillouin zone was sampled using a $12\times 12 \times 16$ Monkhorst-Pack k-point grid~\cite{28} for self-consistent calculations. To obtain a stable equilibrium configuration, full structural relaxation was performed, where both atomic positions and cell parameters were optimized by minimizing the atomic forces. The force convergence threshold was set to 10$^{-3}$ eV/$\text{\AA}$ per ion, ensuring accurate equilibrium geometries before computing the electronic and vibrational properties. Following structural relaxation, the phonon spectra were computed within density functional perturbation theory (DFPT).

\section{Results and Discussion}
The room temperature XRD pattern for Ta$_{0.2}$Nb$_{0.2}$V$_{0.2}$Ti$_{0.2}$X$_{0.2}$ (X = Zr and Hf) are presented in Figure 1 (a) and (b). Le Bail of the XRD data confirms that the material crystallizes in a cubic bcc structure (space group: Im-3m). The obtained lattice parameters are a = b = c = 3.305 (1) and 3.307 (1)  \text{\AA} for X = Zr and Hf, respectively. The microstructure and chemical homogeneity were examined using Field Emission Scanning Electron Microscopy (FE-SEM) and Energy Dispersive Spectroscopy (EDS). The EDS mappings (Figure 1 (c and d)), conducted at ambient temperature, reveal a stoichiometric chemical composition (Ta$_{0.2}$Nb$_{0.2}$V$_{0.2}$Ti$_{0.2}$Hf$_{0.2}$) with an excellent uniform distribution of all elements across the sample on a micrometer scale for X = Hf sample. Multiple regions of the sample were analyzed to confirm the consistency of these observations. For the X = Zr sample, a little deviation from the equiatomic ratio (Ta$_{0.18}$Nb$_{0.20}$V$_{0.17}$Ti$_{0.22}$Zr$_{0.23}$) is observed. A Vickers microhardness of 417 HV and 469 HV (Figure S1 and S2 in the supporting information file, \cite{29}) is observed for X = Zr and Hf, respectively, and these are significantly higher than that of the similar HEA superconductors \cite{30, 31}.

\begin{figure*}
\includegraphics[width=2.0\columnwidth]{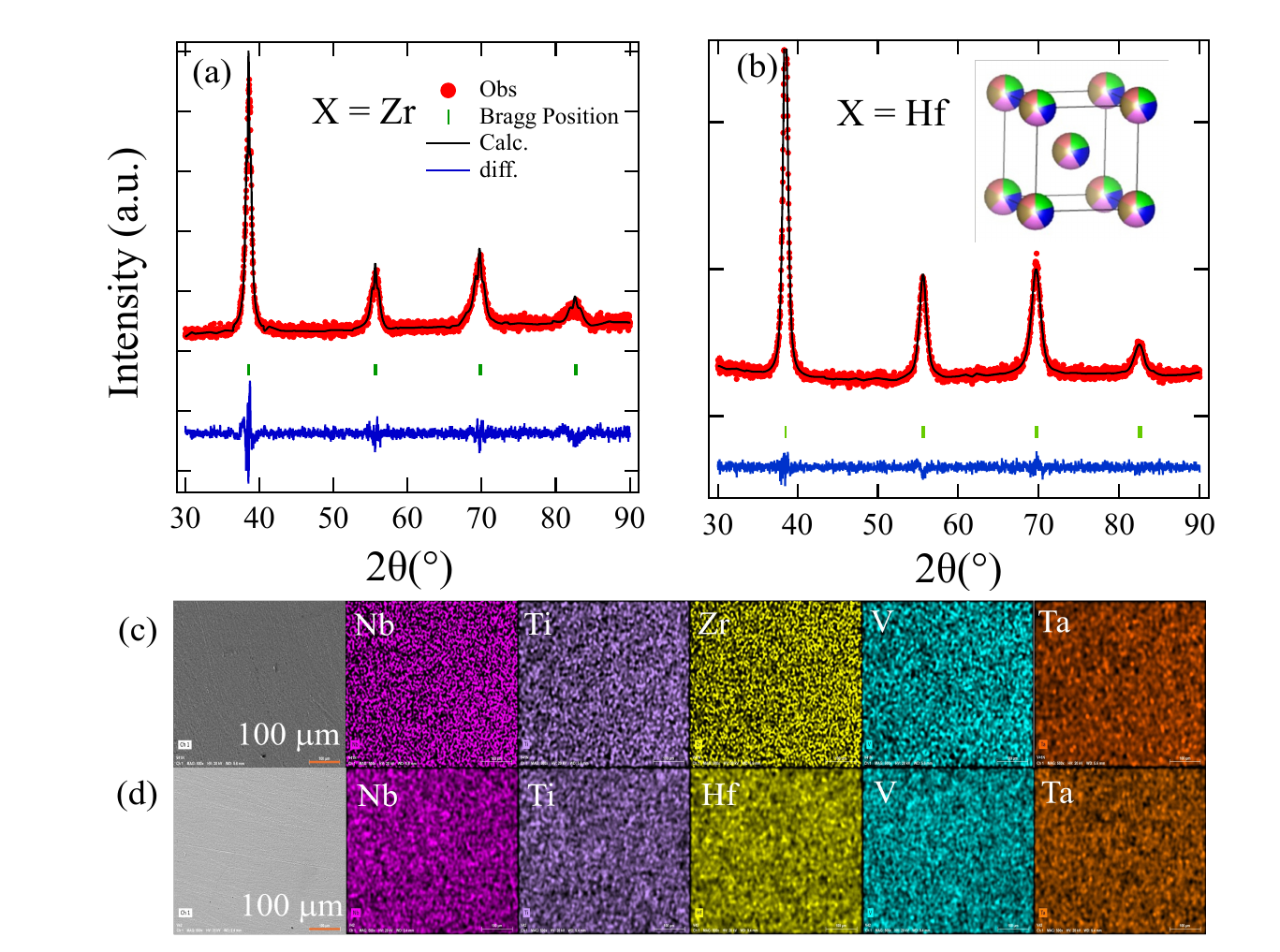}
\caption{\label{Fig1:xrd}(a), (b) Le Bail plots of the X-ray diffraction (XRD) patterns obtained at room temperature; (c) and (d) room-temperature FE-SEM image along with EDS elemental mapping for X = Zr and Hf samples.}
\end{figure*}

\begin{figure*}[t]
\includegraphics[width=2.0\columnwidth]{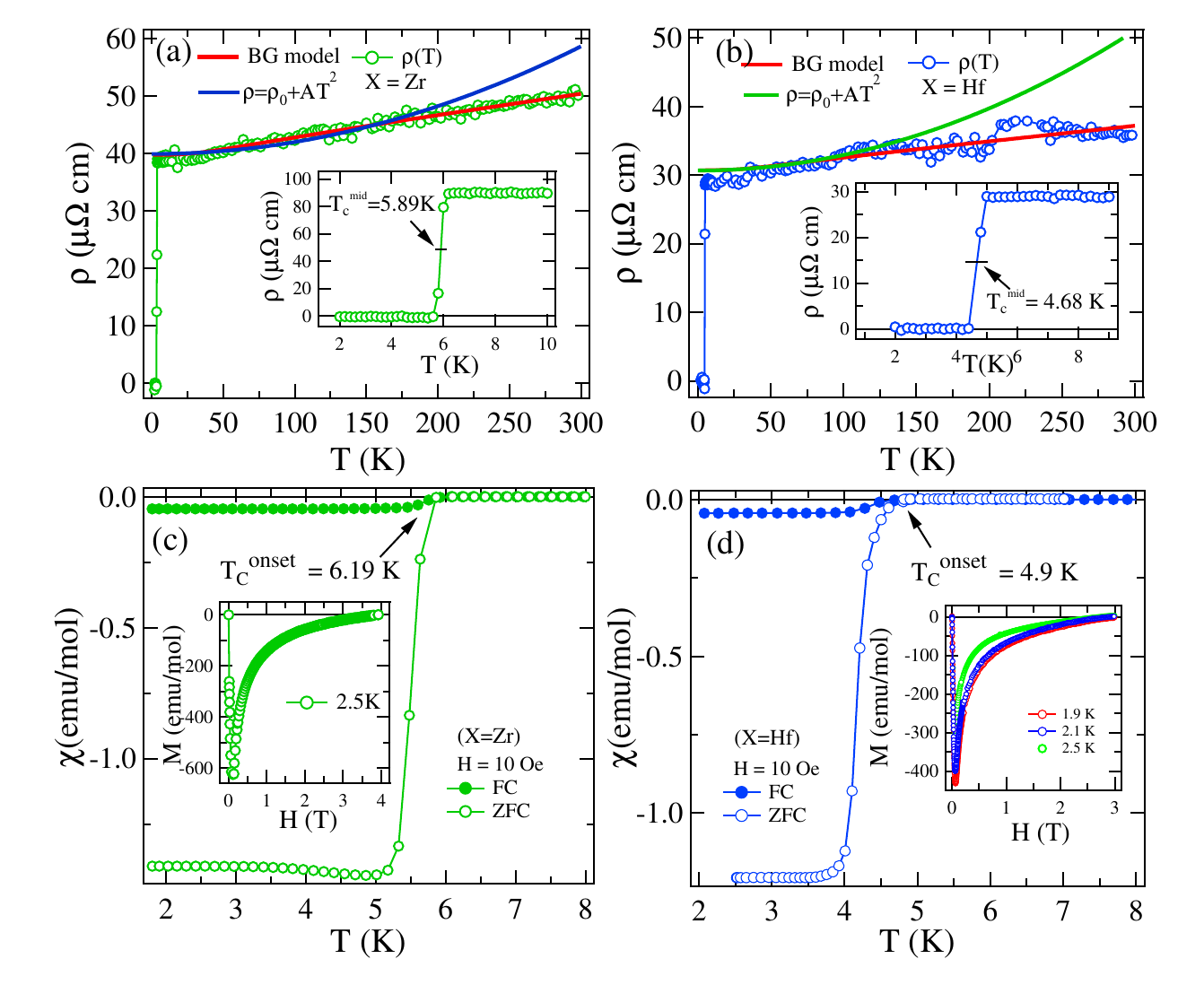}
\caption{\label{Fig2:resis}(a), (b) Temperature-dependent electrical resistivity measured in the absence of magnetic field; and (c), (d) temperature-dependent dc-magnetic susceptibility measured at 10 Oe for X = Zr and Hf samples. Insets in figure (c) and (d) illustrate the variation of magnetization with magnetic field at different fixed temperatures.}
\end{figure*}

\begin{figure*}
\includegraphics[width=2.0\columnwidth]{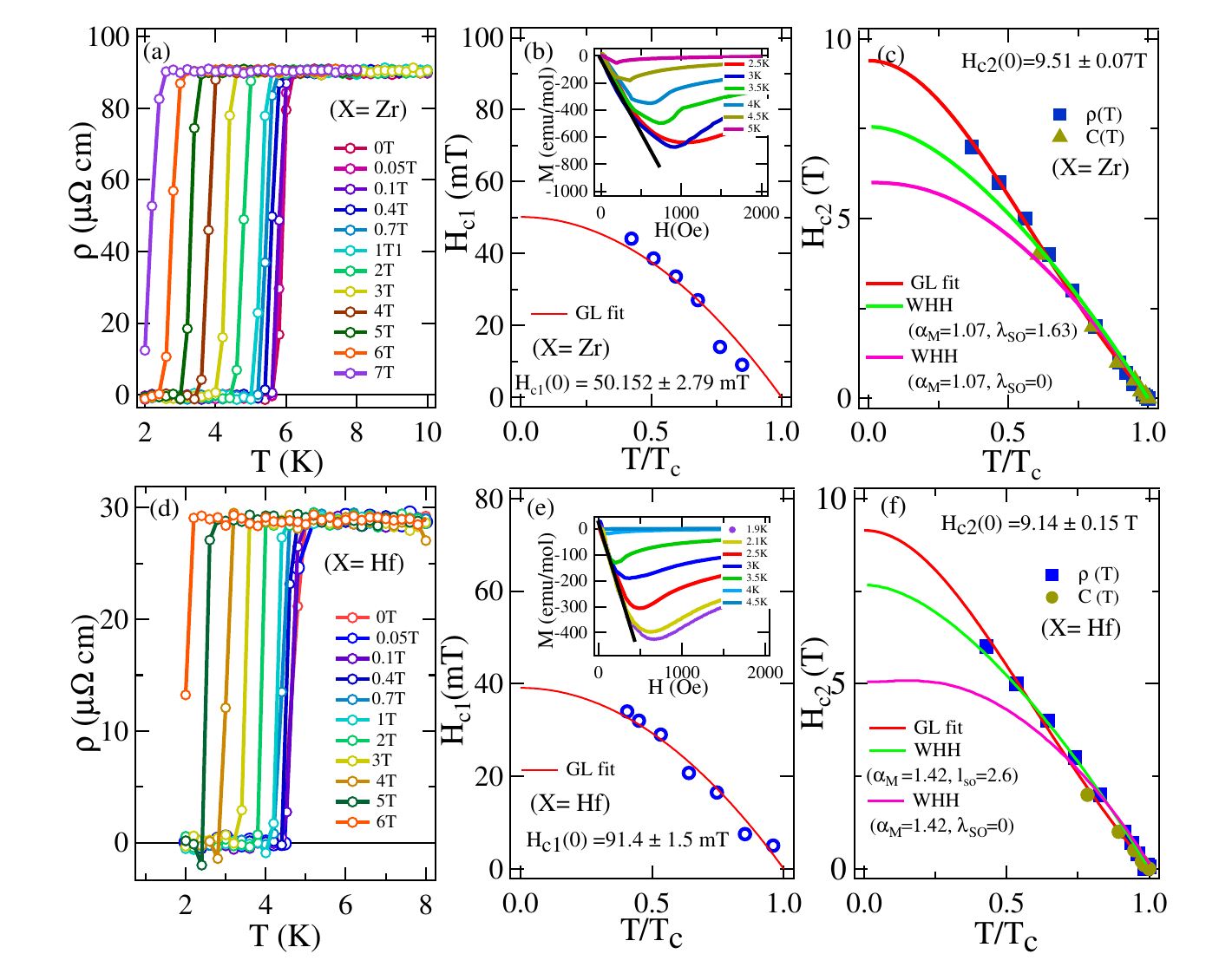}
\caption{\label{Fig3:SPC} (a), (d) Resistivity as a function of temperature under different magnetic fields for X = Zr and Hf samples, respectively; (b),(e) temperature dependence of the lower critical field (H$_{c1}(0)$) and (e),(f) upper critical field (H$_{c2}(0)$) for X = Zr and Hf, respectively. Inset in figures (b) and (e) illustrate M vs. H plots at different fixed temperatures. The upper critical field (H$_{c2}(0)$) is analyzed with the GL and WHH model \cite{32, 33}.}
\end{figure*}

\begin{figure*}
\includegraphics[width=2.0\columnwidth]{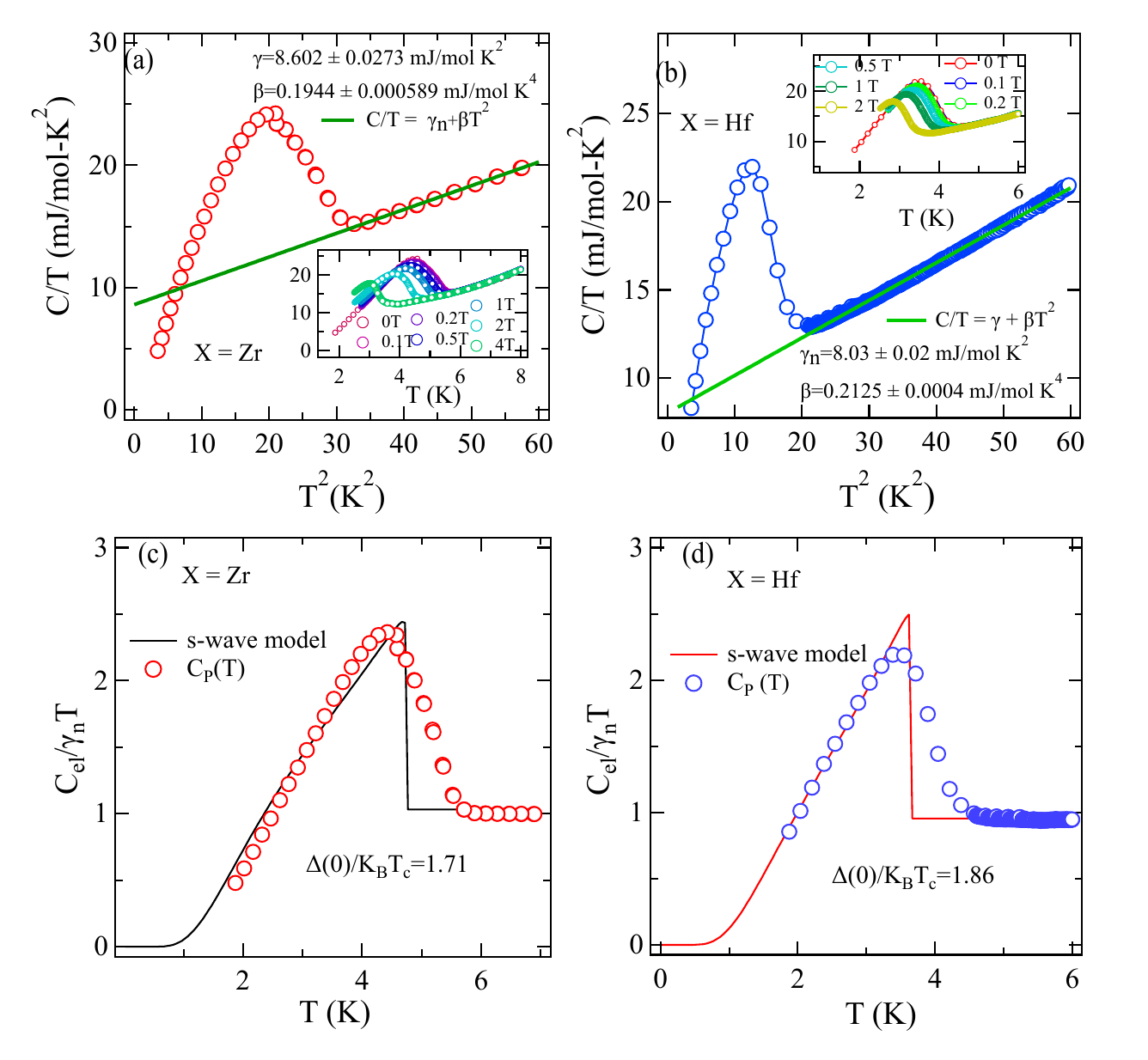}
\caption{\label{Fig4:Phase} (a), (b) C/T vs T$^2$ plots for X = Zr and Hf samples, measurement was performed with zero magnetic field. The normal-state-specific heat data at low temperatures can be described using the Debye model. The inset in the figures highlights the C/T vs T plot measured under different applied magnetic fields. (c), (d) The temperature variation of the electronic specific heat for X = Zr and Hf, respectively.   The electronic-specific heat is modeled using the BCS s-wave equation.}
\end{figure*}
The temperature variation of the electrical resistivity, $\rho$(T), for X = Zr and Hf, measured in the absence of an applied magnetic field, are presented in Figure 2 (a, b). In the low-temperature region, the resistivity begins to drop at $T_c^{\text{onset}}$ = 6.19 K and 5 K and reaches zero at $T_c^{\text{zero}}$ = 5.9 and 4.68 K for X = Zr and Hf respectively, indicating the transition to a superconducting state. The resistivity above the superconducting onset temperature increases slowly with temperature indicating poor metallic behavior. We have analyzed the normal state resistivity data for both materials with the Bloch-Grueneisen (BG) model, as written below (Figure 1 (a and b)).
\begin{equation}
\rho(T) = \rho_0 + C \left(\frac{T}{\theta_D}\right)^n \int_0^{\theta_D/T} \frac{x^n}{(e^x - 1)(1 - e^{-x})} \, dx
\end{equation}
In this equation (1), $\rho_0$ is the residual resistivity, and the second term is BG expression \cite{34} in which C depends on the intrinsic properties of the material, n depends on the electron-phonon scattering and $\theta_D$ is Debye temperature. The expression is best fitted using n = 3 for both the samples, and yields $\rho_0$ = 39.5 and 30.7 $\mu$ $\Omega$ $\text{cm}$ and $\theta_D$ = 215 $\text{K}$ and 211 $\text{K}$ for X = Zr and Hf respectively. We have also determined the Kadowaki-Woods ratio (\(R_{KW} = \frac{A}{\gamma_n^2}\)), which quantifies the strength of electron-electron interactions. The coefficient \(A\) reflects the low-temperature electron-electron scattering and can be derived by fitting the normal state resistivity to the power law expression \(\rho = \rho_0 + AT^2\), as shown in Figure 2 (a) and (b). From this fitting, we obtained \(A = 2.15(3) \times 10^{-4}\) and \(2.3(3) \times 10^{-4}\ \mu\Omega\text{-cm K}^{-2}\) for X = Zr and Hf respectively. Considering $\gamma_n = 8.60 (3)$ (for X = Zr, taken from the specific heat fitting, Figure 4) and \(8.03 (2)\, \text{mJ mol}^{-1} \text{K}^{-2}\) (for X = Hf), the calculated \(R_{KW}\) is $ 0.29 \times 10^{-5}$ and \(0.36 \times 10^{-5} \, \mu\Omega\text{-cm K}^2\text{mJ}^{-2}\text{mol}^2\) for X = Zr and Hf respectively. The values of Kadowaki-Woods ratio, being less than \(1 \times 10^{-5} \, \mu\Omega\text{-cm K}^2\text{mJ}^{-2}\text{mol}^2\), suggests that both the materials are weakly correlated.\\

Figures 2(c) and (d) illustrate the temperature dependence of magnetic susceptibility, $\chi(T)$, measured under a 10 Oe applied magnetic field in both ZFC and FC modes. The ZFC data show a sharp diamagnetic response below $T_c^{\text{onset}} = 6.0$ K and 4.9 K for X = Zr and Hf, respectively, marking the onset of superconductivity for both these samples. The previously reported magnetic susceptibility data for the X = Hf sample indicate a similar superconducting transition temperature ($T_c^{\text{onset}} = 4.8$ K) \cite{35}. However, their resistivity measurements show a higher $T_c^{\text{onset}}$ of approximately 5.5 K. In contrast, our measurements show consistent superconducting transition temperatures ($T_c^{\text{onset}} \approx 4.9$ K) in resistivity, magnetization, and specific heat data. The difference in the transition temperature observed in different measurements could arise from the microscopic inhomogeneity in the sample. Furthermore, the magnetic hysteresis curve, $M(H)$, measured at different fixed temperatures, exhibits the typical behavior of a type-II superconductor (inset in Figure 2 (c) and (d)). The lower critical field ($H_{c1}(0)$) is determined using the points where the $M(H)$ curve deviates from its initial linearity and is estimated to be 50 mT and 38 mT for X = Zr and Hf at zero temperature using the Ginzburg-Landau approximation:
\begin{equation}
H_{c1}(T)=H_{c1}(0)\left[1-\left(\frac{T}{T_{c}}\right)^{2}\right]
\label{eqn1:Hc}
\end{equation}

The upper critical field, (H$_{c2}(0)$), was determined through a series of measurements, including temperature-dependent electrical resistivity ($\rho$(T)) (Figure 3 (a) and (d)) and specific heat (Figure 4), measurements at different fixed magnetic fields. In the resistivity measurements, the midpoint of the resistivity drop is considered as T$_C$ to calculate the H$_{c2}(0)$. As the magnetic field is increased, resistivity, dc-magnetization, and specific heat measurements show that $T_c$ values shift to lower temperatures. The H$_{c2}(T)$ vs. T curves obtained from these measurements are shown in Figure 3 ((c) and (f)). The H$_{c2}(T)$ vs. T behaviour for both the samples is analyzed using the phenomenological Ginzburg-Landau (GL) model:

\begin{equation}
H_{c2}(T) = H_{c2}(0) \frac{1-(T/T_c)^2}{1+(T/T_c)^2}
\end{equation}

The GL model fits the \( H_{c2}(T) \) data accurately and gives \( H_{c2}(0) = 9.51 \pm 0.07 \, \text{T} \) and 9.14 $\pm$ 0.15 T for X = Zr and Hf respectively.
Interestingly, for the X = Hf sample, the obtained value of the H$_{c2}(0)$ is higher than the Pauli paramagnetic field (1.85T$_C$), suggesting the possibility of unconventional superconductivity in this sample. Surpassing the Pauli limiting field is extremely rare in high entropy alloy superconductors. For instance, a large H$_{c2}(0)$ is observed in a magnetic cation Cr containing noncentrosymmetric high entropy alloy \cite{16} and recently studied ScVTiHfNb \cite{15} (T$_C$ = 4.17) shows a high upper critical field surpassing the Pauli paramagnetic limit. In addition to the potential for unconventional superconductivity (either through unique pairing mechanisms or superconducting state symmetries), these materials exhibit enhanced stability in high magnetic fields (exceeding the Pauli limit). This makes them particularly valuable for high-field applications, such as MRI machines, particle accelerators, and magnetic confinement fusion systems. The presence of Ta and Hf in the present sample introduces spin-orbit coupling (SOC) and possibly reduces the influence of the Pauli paramagnetic effect, allowing H$_{c2}(0)$ to surpass the Pauli limiting field. However, compared to Hf, since Zr has aed SOC effecteffect,s lower than the Pauli paramagnetic field (1.85T$_C$) in the sample the sample X = Zr.ype-II superconductors, superconductivity (pair breaking) can be disrupted by two primary mechanisms: the orbital limiting effect and the Pauli paramagnetic effect. To investigate the mechanism of pair breaking in these materials, H$_{c2}(0)$ is also examined using the Werthamer-Helfand-Hohenberg (WHH) model \cite{32, 33}, which accounts for the effects of Pauli spin paramagnetism ($\alpha$, also Maki parameters \cite{36, 37}) and spin-orbit scattering ($\lambda_{SO}$). The Maki parameter \cite{37}, which also estimates the relative strengths of the orbital and Pauli-limiting fields, is given by:
\begin{equation}
\alpha_M = \sqrt{2} \frac{H_{c2}^{\text{orb}}(0)}{H_{c2}^{p}(0)}
\label{eqn1:maki}
\end{equation}
In the dirty limit, for type-II superconductor, the orbital limit of the upper critical field \( H_{c2}^{\text{orb}}(0) \) is expressed by:
\begin{equation}
H_{c2}^{\text{orb}}(0) = -0.693 T_c \left. \frac{dH_{c2}(T)}{dT} \right|_{T=T_c}
\label{eqn1:orbit}
\end{equation}
An initial slope of \( -2.04 \pm 0.02 \, \text{T K}^{-1} \) and \( -2.68 \pm 0.02 \, \text{T K}^{-1} \) near \( T_C \) yields \( H_{c2}^{\text{orb}}(0) = 8.32 \, \text{T} \) and 8.72 T for X = Zr and Hf respectively. The Pauli-limiting field is given by \( H_{c2}^{p}(0) = 1.85 T_c = 10.9 \, \text{T} \) and 8.69 T for X = Zr and Hf sample, respectively. The fact that H$_{c2}^{p}(0)$ is smaller than \(H_{c2}^{\text{orb}}(0)\) for the X = Hf sample indicates that \(\mu_0 H_{c2}\) at low temperatures is limited by the Pauli spin susceptibility rather than the conventional pair-breaking (orbital) effect. This observation points to an anomalous property in X = Hf sample. The Maki parameter is calculated to be \( \alpha_M = 1.07 \) and 1.42 for X = Zr and Hf samples, respectively. By setting these values of $\alpha$ and varying the $\lambda_{SO}$, we have fitted our data using the WHH model \cite{32, 33, 38}:

\begin{align}
\ln \left(\frac{1}{t}\right) = \left(\frac{1}{2} + \frac{i \lambda_{SO}}{4 \gamma}\right) \psi\left(\frac{1}{2} + \frac{\bar{h} + \lambda_{SO}}{2} + \frac{i \gamma}{2t}\right)\\
+ \left(\frac{1}{2} - \frac{i \lambda_{SO}}{4 \gamma}\right) \psi\left(\frac{1}{2} + \frac{\bar{h} + \lambda_{SO}}{2} - \frac{i \gamma}{2t}\right) - \psi\left(\frac{1}{2}\right)
\label{eqn3:WHH}
\end{align}

where:
\[
\gamma \equiv \left(\alpha \bar{h}\right)^2 - \left(\frac{\lambda_{SO}}{2}\right)^2,
\]
and
\[
h^* \equiv \left. \frac{d\bar{h}}{dt} \right|_{t=1} = \frac{\pi^2 \bar{h}}{4} = \left. \frac{dH_{c2}}{dt} \right|_{t=1}
\]
with \( t = \frac{T}{T_c} \) is the reduced temperature, \( \bar{h} \) is the reduced field, \( \lambda_{SO} \) is the spin-orbit coupling parameter, \( \alpha \) is the Maki parameter, \( \gamma \) is a parameter combining the effects of Pauli paramagnetism and spin-orbit coupling, \( \psi \) is the digamma function. In this formula: \( h^* \) represents the initial slope of the reduced field \( \bar{h} \) with respect to the reduced temperature \( t \) at \( t = 1 \) (i.e., near \( T_c \)). 

The fitting is shown in Figure 3 ((c) and (f)), and it provides $\lambda_{SO}$ = 1.63 and 2.6 for X = Zr and Hf samples, respectively. In conventional BCS superconductors, where \(\alpha\) and \(\lambda_{\text{SO}}\) are typically zero, the upper critical field remains well below the Pauli limiting field \cite{33}. Nonetheless, there are several systems where a large \(\mu_0 H_{c2}(0)\) and deviations from the WHH model is observed. The obvious examples are noncentrosymmetric superconductors \cite{39}, iron-based high-temperature superconductors \cite{40}, and systems with strong spin-orbit scattering \cite{41}. We suggest that spin-orbit scattering is a key factor in enhancing the upper critical field in the Hf containing sample, particularly due to the significant value of \(\lambda_{\text{SO}}\) in this system. 

We have utilized the values of \(H_{c2}(0)\) to estimate the Ginzburg-Landau coherence length, \(\xi_{\mathrm{GL}}(0)\), via the following expression \cite{42}:
\begin{equation}
    H_{c2}(0) = \frac{\Phi_{0}}{2\pi\xi_\mathrm{GL}^{2}}
\label{eqn3:xi}
\end{equation}
where \(\Phi_{0} = 2.07 \times 10^{-15} \, \text{Tm}^2\) represents the magnetic flux quantum. By substituting the H$_{c2}(0)$ values, we find a similar coherence length for both the samples (\(\xi_{\mathrm{GL}}(0) = 5.8\, \text{nm}\) and 6 nm for X = Zr and Hf, respectively). The magnetic penetration depth, \(\lambda_{\mathrm{GL}}(0)\), is determined using the \(H_{c1}(0)\) and the previously calculated \(\xi_{\mathrm{GL}}(0)\) from the following relation \cite{43}:
\begin{equation}
H_{c1}(0) = \frac{\Phi_{0}}{4\pi\lambda_\mathrm{GL}^2(0)}\left(\ln\frac{\lambda_\mathrm{GL}(0)}{\xi_\mathrm{GL}(0)}+0.12\right)
\label{eqn4:lambda}
\end{equation}
Substituting \(H_{c1}(0)\) and \(\xi_{\mathrm{GL}}(0)\) values, the penetration depth \(\lambda_{\mathrm{GL}}(0)\) is calculated to be 981.9 \text{\AA} and 1164 \text{\AA} for X = Zr and Hf, respectively. The Ginzburg-Landau parameter, \(\kappa_{\mathrm{GL}}\), which characterizes the type of a superconductor, is given by \(\kappa_{\mathrm{GL}} = \frac{\lambda_{\mathrm{GL}}(0)}{\xi_{\mathrm{GL}}(0)}\). The calculated value of \(\kappa_{\mathrm{GL}}(0) = 16.7\) and 19.38 for X = Zr and Hf, respectively, and these are significantly greater than \(\frac{1}{\sqrt{2}}\), indicating the type-II superconductivity in both the materials.

To confirm the bulk superconducting nature of the material, we have conducted temperature variation of the specific heat measurements, \( C (T) \) for both materials. Specific heat data are collected under applied magnetic fields ranging from 0 to 2 T. As shown in Figure 4 (a) and (b), the specific heat for both samples exhibits a distinct anomaly that shifts to lower temperatures with an increase in the field strength, indicating the emergence of the superconducting transition. The specific heat data in the normal state were well-fitted using the following expression:
\begin{equation}  
\frac{C}{T} = \gamma_n + \beta T^{2},
\label{eqn6:SH}    
\end{equation} 
Sommerfeld coefficient (\(\gamma_n = 8.60(2) \, \text{mJmol}^{-1}\text{K}^{-2}\) and 8.03 (2) $\text{mJmol}^{-1}\text{K}^{-2}$ for X = Zr and Hf respectively) represents the electronic part, and $\beta$ (= 0.2125(4) and 0.1944 (5) $\text{mJmol}^{-1}\text{K}^{-4}$ for X = Zr and Hf) corresponds to the lattice part to the specific heat. These parameters are essential for calculating the density of states at the Fermi level \([D_c(E_\mathrm{F})]\), the Debye temperature \((\theta_D)\), and the electron-phonon coupling constant \((\lambda_{e-ph})\). The Debye temperature \(\theta_D\) is determined as 215.4 (1) K and 209.1 (1) K for X = Zr and Hf samples, respectively, and is related to \(\beta\) by the expression \cite{44}:
\begin{equation}
    \theta_D = \left(\frac{12\pi^{4}RN}{5\beta}\right)^{\frac{1}{3}}
    \label{eqn8:thetaD}
\end{equation}
These \(\theta_D\) values matched well with the obtained values from the BG fitting of the normal state resistivity data for both samples.

To investigate the pairing symmetry and superconducting gap ratio, we have analyzed the electronic specific heat, \(C_\text{el}\). Figures 4(c) and 4(d) display the electronic part (\(C_{el}\)) to the specific heat obtained by subtracting the phononic part for X = Zr and Hf samples, respectively. The following expression describes the superconducting state normalized entropy for a single-gap BCS superconductor \cite{15, 38}:
\begin{equation}
\frac{S}{\gamma_{n}T_{c}} = -\frac{6}{\pi^2}\left(\frac{\Delta(0)}{k_{B}T_{c}}\right)\int_{0}^{\infty}\left[ f\ln(f)+(1-f)\ln(1-f)\right]dy
\label{eqn10:BCS1}
\end{equation}
where Fermi-Dirac distribution function = \( f(\xi) = \left[\exp\left(E(\xi)/k_{B}T\right)+1\right]^{-1} \), energy of quasiparticles relative to the Fermi energy = \( E(\xi) = \sqrt{\xi^{2}+\Delta^{2}(t)} \), \( y = \xi/\Delta(0) \), \( t = T/T_{c} \), and \(\Delta(t)\) = temperature-dependent superconducting gap \( = \tanh\left[1.82(1.018((1/t)-1))^{0.51}\right]\).
\begin{table}[h!]
\caption{Superconducting state parameters of Ta$_{0.2}$Nb$_{0.2}$V$_{0.2}$Ti$_{0.2}$X$_{0.2}$.}
\label{superconducting properties}
\begin{center}
\begin{tabular*}{1.0\columnwidth}{l@{\extracolsep{\fill}}llll}\hline\hline
Properties& unit& X = Hf& X = Zr\\
\hline
\\[0.5ex]                                  
$T_c^{onset}$& K& 5.0 $\pm$ 0.2& 6.2 $\pm$ 0.2\\ 
$T_{c,mid}$& K& 4.7 $\pm$ 0.1& 5.9 $\pm$ 0.1\\
$H_{c1}(0)$& mT& 37.54 $\pm$ 2.01& 50.5 $\pm$ 2.0\\                     $H_{c2}(0)$& T& 9.14 $\pm$ 0.15& 9.51 $\pm$ 0.07\\
$H_{c2}^{P}(0)$& T& 8.63& 10.92\\
$\xi_{GL}$& \text{\AA}& 60 $\pm$ 1& 58.8 $\pm$ 1.0\\
$\lambda_{GL}$& \text{\AA}& 1164 $\pm$ 11& 981 $\pm$ 10\\
$\kappa$& &19.38 $\pm$ 0.08& 16.7 $\pm$ 0.1\\
$\gamma$& mJmol$^{-1}$K$^{-2}$& 8.03 $\pm$ 0.02& 8.60 $\pm$ 0.02\\
$\beta$ & mJmol$^{-1}$K$^{-4}$& 0.2125 $\pm$ 0.0004& 0.1944 $\pm$ 0.0004\\
$\theta_{D}$& K& 209.1 & 215.4 \\
$\omega_{ln}$& K& 200.27 & 179 \\
$\lambda_{e-ph}$&  &0.63 $\pm$ 0.02& 0.76 $\pm$ 0.02\\
$\Delta(0)/k_{B}T_{C}$(eq. 11)&   &1.86 $\pm$ 0.01& 1.71 $\pm$ 0.01\\
$\Delta(0)/k_{B}T_{C}$(eq. 14)&   &1.80 $\pm$ 0.01& 1.82 $\pm$ 0.01\\
$\Delta C_{el}/\gamma T_{C}$&   &1.54 $\pm$ 0.01& 1.62 $\pm$ 0.01
\\[0.5ex]
\hline\hline
\end{tabular*}
\par\medskip\footnotesize
\end{center}
\end{table}

\begin{figure*}
\includegraphics[width=2.0\columnwidth]{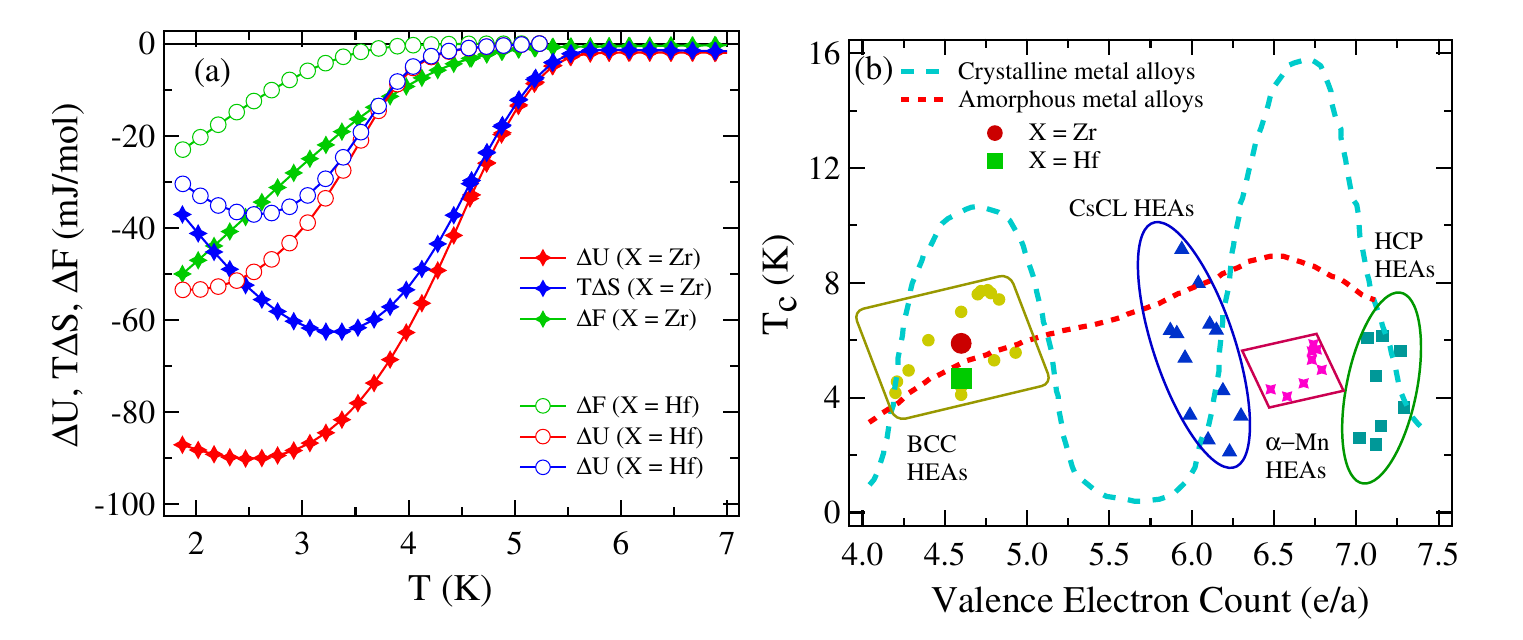}
\caption{\label{Fig5:Calc} (a) The temperature dependence of change in internal energy (\(\Delta U\)), change in entropy (multiplied by temperature, \(T \Delta S\) and change in free energy (\(\Delta F\)) for X = Zr and Hf samples. (b) The superconducting transition temperature vs. the valence electron concentration (VEC) plot. Superconducting transition temperature and VEC for various crystalline metals \cite{45}, amorphous metals \cite{46}, and different cubic HEA alloys are compared here \cite{7, 15, 47}.}
\end{figure*}

\begin{figure*}
\includegraphics[width=2.0\columnwidth]{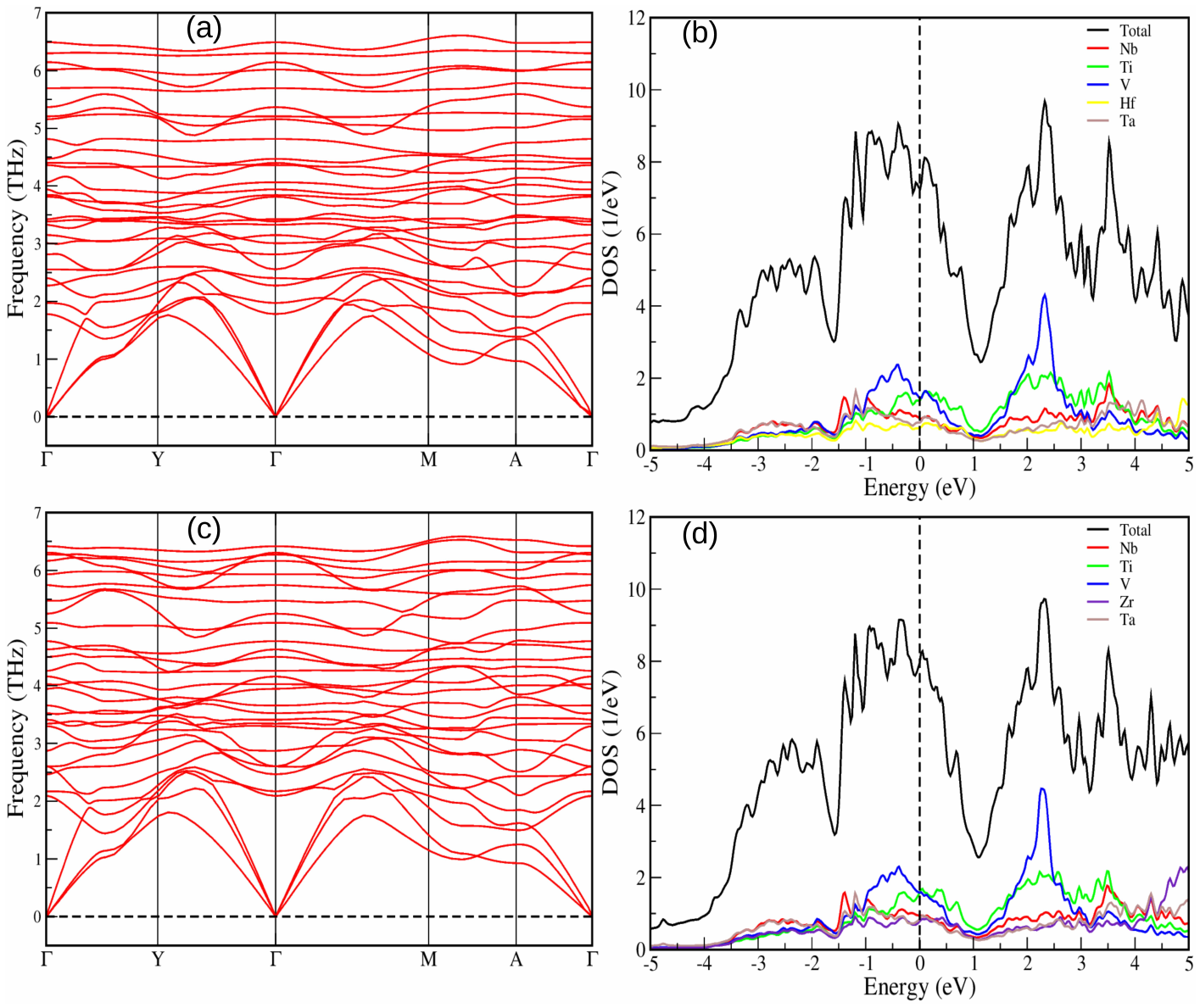}
\caption{\label{Fig6:DOS} Phonon dispersion relations along high-symmetry directions for Ta$_{0.2}$Nb$_{0.2}$V$_{0.2}$Ti$_{0.2}$X$_{0.2}$ with X = Hf and Zr are shown in panels (a) and (c), respectively. The total and partial density of states are presented in panel (b) for X = Hf and in panel (d) for X = Zr.}
\end{figure*}

The first derivative of the normalized entropy with respect to temperature is related to the normalized electronic specific heat below \(T_c\) by:
\begin{equation}
\frac{C_{el}}{\gamma_{n}T_{c}} = t\frac{d(S/\gamma_{n}T_{c})}{dt}
\label{eqn11:BCS2}
\end{equation}

For X = Zr sample, the obtained superconducting gap ratio \(\alpha_{BCS} = \Delta(0)/k_{B}T_{C} = 1.71\), which is close to the BCS value of \(\alpha_{BCS} = 1.76\). However, the BCS fitting estimates a lower value of the specific heat jump (\(\Delta C_{el}/\gamma_nT_{C}\) = 1.41). Our estimated \(\Delta C_{el}/\gamma_nT_{C}\) = 1.62 as shown in the Supplementary Information File (Figure S3, \cite{29}). The disorder-induced broad transition (in specific heat) could be the reason for this discrepancy. By fitting this model with the data for the sample X = Hf, we obtain a superconducting gap ratio \(\alpha_{BCS} = \Delta(0)/k_{B}T_{C} = 1.86\) (\(\Delta C_{el}/\gamma_nT_{C}\) = 1.54), which is significantly higher than the BCS value of \(\alpha_{BCS} = 1.76\), and also exceeds the value reported for the equimolar ScVTiHfNb (showing a high critical field) \cite{15}. 

The specific heat data of these samples are further analyzed in the strong coupling limit using the following equations by considering $\frac{T_C}{\omega_{ln}} << 1$.

\begin{equation}
    \frac{\Delta C_{el}}{\gamma \ T_C} = 1.43 \left[ 1 + 53 \left( \frac{T_C}{\omega_{ln}} \right)^2 \ln \left( \frac{\omega_{ln}}{3 \ T_C} \right) \right]
    \label{phononFrequency}
\end{equation}
\begin{equation}
    \frac{2\Delta (0)}{k_B \ T_C} = 3.53 \left[1 + 12.5 \left( \frac{T_C}{\omega_{ln}} \right)^2 \ln \left( \frac{\omega_{ln}}{2 \ T_C} \right) \right]
    \label{BCSgap}
\end{equation}

The value of the specific heat jump (1.62 and 1.54 for X = Zr and Hf respectively) is used to estimate the logarithmically averaged phonon frequency ($\omega _{ln}$) for both the samples. The estimated $\omega _{ln}$ = 179 K for X = Zr and 200.27 K for X = Hf. These values are comparable to the Debye temperature. The calculated value of the \( \Delta(0)/k_{B}T_{C}\) is 1.82 for X = Zr and 1.80 for X = Hf sample.

To quantitatively assess the potential for superconductivity, we computed the electron-phonon coupling parameter, $\lambda$ from our DFT calculations. It is expressed by the following equation:
\begin{equation}
    \lambda = \sum_{q\nu} \lambda_{q\nu},
\end{equation}
where $\lambda_{q\nu}$ is the electron phonon coupling strength associated with a specific phonon mode $\nu$ and wave-vector $q$. The $\lambda_{q\nu}$ is given by:
\begin{equation*}
\begin{split}
\lambda_{q\nu} = \frac{\gamma_{q\nu}}{\pi N(\varepsilon_F)}\omega_{q\nu}^2
\end{split}
\end{equation*}
where, $\gamma_{q\nu}$ is phonon linewidth, N($E_F$) is the density of states (DOS) at the Fermi energy and $\omega$ is the frequency associated with specific phonon modes.
From our calculations we obtained an electron-phonon coupling constant of $\lambda = 0.63$ for $X$ = Hf and $\lambda = 0.76$, for $X$ = Zr. These values fall in the moderately-coupling regime. The computed values of the \(\lambda\) can be used to estimate the value of the screened coulomb repulsion constant ($\mu^*$) using the Allen-Dynes formula \cite{48}: 
\begin{equation}
T_C = \frac{\omega_{\ln}}{1.2} \exp \left( -\frac{1.04(1+ \lambda)}{\lambda - \mu^* (1+0.62 \ \lambda)} \right)
\label{AllenDynes}
\end{equation}
The obtained values of $\mu^* = 0.13$ for Zr and 0.112 for Hf are in a good agreement with commonly used for weak and moderately coupled intermetallic superconductors $\mu^* = 0.13$ \cite{15, 49}. Therefore, electronic-specific heat measurements, combined with the electron-phonon coupling constant, suggest that both samples exhibit moderately coupled BCS superconductivity \cite{50}. A summary of all the obtained parameters is provided in Table 1. 

In general, the behavior of specific heat of a superconducting material is different in the superconducting state and the normal state. The electronic part of the specific heat data (\(C_{el}\)) gives a direct manifestation of the change in internal energy (\(\Delta U\)), entropy (\(\Delta S\)), and the free energy (\(\Delta F\)) of the material \cite{51}. The difference in the internal energy of the material can be estimated using the following relation:
\begin{equation}
\Delta U(T) =  \int_T^{0} (C_{el}(T')-\gamma_n T') \, dT'
\end{equation}
where the symbols have been explained previously and $T'$ = temperature. Similarly, the change in entropy of the system can be calculated by the following equation:
\begin{equation}
\Delta S(T) = \int_T^{0} \frac{(C_{el}(T')-\gamma_n T')}{T'}  \, dT'
\end{equation}
These calculated parameters, \(\Delta U\) and \(\Delta S\)  help us to understand the thermodynamics of the superconductors in relation to the difference in free energy between the superconducting and normal states \cite{52} with the following equation:
\begin{equation}
\Delta F(T) = \Delta U(T) - T \Delta S(T)
\end{equation}
It is observed from Figure 5 (a) that the values of \(\Delta F\) are negative for both samples below the superconducting transition temperature. In BCS framework, the negative value of \(\Delta F\) implies enhanced stability of the superconducting state due to the formation of cooper pairs. As the temperature rises, the entropy of the material become significantly large and the gradual breaking of cooper pairs takes place. This implies the quasiparticles start to break out of the superconducting state. Above the superconducting transition temperature, the \(\Delta F\) is still stabilized, but the system shows a transition to the normal metallic state. 
Figure 5 (b) presents the relationship between the critical temperature \(T_C\) and the valence electron concentration (VEC) for both the samples. To compare, data from both crystalline and amorphous metals are included \cite{45, 46} in the graph. The data of other HEA superconductors are also included in the figure \cite{15, 21, 47}. The \(T_C\) dependence on VEC for HEA superconductors is expected to lie between that of crystalline and amorphous alloys. Therefore, our data is consistent with other HEA superconductors with similar VEC values.

To further provide crucial insights into the structural stability and electronic structure of Ta$_{0.2}$Nb$_{0.2}$V$_{0.2}$Ti$_{0.2}$X$_{0.2}$ (X = Hf and Zr), we presented our results of DFT based calculations in Figure 6. The phonon dispersion spectrum (Fig. 6 (a) and (c)) reveals the absence of imaginary frequencies across the Brillouin zone, establishing the dynamical stability of these alloys. This confirms that the atomic configuration obtained after structural relaxation corresponds to a mechanically stable phase. Furthermore, the presence of low-energy optical phonon modes suggests finite electron-phonon interactions. However, the overall phonon bandwidth is relatively large. The phonon modes extend up to approximately 7 THz. These results indicate that the material exhibits a broad range of vibrational excitations.
The computed electronic density of states (DOS) are illustrated in Figs. 6 (b) and (d), respectively). The total DOS for both samples demonstrates the metallic behavior with large spectral weight at Fermi energy ($E_F$). A significant DOS at ($E_F$) suggests a high carrier density, which is a favorable characteristic for electrical conductivity and potential superconducting properties. Further, the partial DOS show that a substantial contribution near the Fermi level arises from Ti and V derived states, indicating their dominant role in electronic transport. These results along with the experimental observations stimulate further investigations of superconductivity and transport phenomena.  This suggests that while the material exhibits metallic conductivity with a high carrier density, the electron-phonon interaction is not sufficiently strong to induce a high superconducting transition temperature (T$_c$). This is consistent with our experimental results. Thus, our theoretical calculations establish that Ta$_{0.2}$Nb$_{0.2}$V$_{0.2}$Ti$_{0.2}$X$_{0.2}$ (X = Hf and Zr) are a structurally stable, metallic system with a high carrier density but moderate electron-phonon coupling, which limits its superconducting transition temperature.

\section{Conclusion}
In conclusion, we have successfully stabilized high-entropy equiatomic alloys Ta$_{0.2}$Nb$_{0.2}$V$_{0.2}$Ti$_{0.2}$X$_{0.2}$ with X = Zr and Hf. The intrinsic maximal disorder of equiatomic high-entropy alloys, in general, provides a valuable opportunity to investigate further and understand the mechanisms driving superconducting pairing in these disordered materials. Le Bail of the RT-XRD pattern confirms that both samples crystallize in a cubic body-centered cubic (bcc) structure (space group: Im-3m) with lattice parameters of a = b = c =  3.305 (1) and 3.307 (1) \text{\AA} for X = Zr and Hf, respectively. FESEM and EDS measurements highlight a uniform equiatomic distribution of all the elements across the sample on a micrometer scale. Through comprehensive magnetization, electrical resistivity, specific heat measurements, and theoretical studies, we have demonstrated that these alloys exhibit bulk superconductivity with an onset temperature of approximately 5 K for X = Hf and 6.19 K for the X = Zr sample. The normal state behavior is explored using the resistivity (Bloch-Grueneisen model, Kadowaki-Woods ratio) and heat capacity studies. The Kadowaki-Woods ratio highlights weakly correlated behavior for both samples. Our detailed analysis of the superconducting properties highlights moderately coupled, isotropic s-wave superconductivity in both samples. Moreover, the upper critical field, estimated from the transport and thermodynamic measurements, surpasses the Pauli paramagnetic limit for the X = Hf sample, suggesting the potential for unconventional superconducting behavior. The spin-orbit coupling (SOC), due to the existence of Ta and Hf in the sample, is expected to favor the unconventional superconducting behavior in the material. In addition, our first-principles calculations confirm the dynamical stability of both high entropy alloys through the absence of imaginary phonon modes. The electronic structure reveals substantial spectral weight at the Fermi level, supporting the excellent metallic ground state of these HEAs. Furthermore, the computed electron-phonon coupling constants fall in the intermediate coupling regime, consistent with the moderately coupled isotropic s-wave superconductivity inferred from experimental measurements. With its excellent metallic properties in the normal state, high micro-hardness, and enhanced upper critical field, these high-entropy alloys emerge as a strong candidate for future superconducting device applications. 

\section{ACKNOWLEDGMENTS}
S. M. acknowledges the ANRF (previously SERB), Government of India, for the (SRG/2021/001993) Start-up Research Grant. Work in Tallinn, Estonia, was supported by the European Regional Development Fund (Awards TK133 and TK134) and the Estonian Research Council (Projects No. PRG4 and No. PRG1702).

\end{document}